\documentclass[runningheads]{llncs}
\usepackage{graphicx}
\usepackage{times}
\usepackage{epsfig}
\usepackage{graphicx}
\usepackage{amssymb}
\usepackage{pifont}
\usepackage{bm}
\usepackage{amsmath}
\usepackage{amssymb}
\usepackage{dsfont}
\newcommand{\app}{\raise.17ex\hbox{$\scriptstyle\sim$}}

\usepackage{xspace}

\makeatletter
\DeclareRobustCommand\onedot{\futurelet\@let@token\@onedot}
\def\@onedot{\ifx\@let@token.\else.\null\fi\xspace}

\def\eg{\emph{e.g}\onedot} 
\def\ie{\emph{i.e}\onedot}

\def\etal{\emph{et al}\onedot}
\makeatother

\begin{document}
\title{Multi-Scale Attentional Network for \\Multi-Focal Segmentation of Active Bleed \\ after Pelvic Fractures}

\titlerunning{MSAN for Multi-Focal Segmentation of Active Bleed after Pelvic Fractures}
\author{Yuyin Zhou\textsuperscript{1}, David Dreizin\textsuperscript{2}, Yingwei Li\textsuperscript{1}, 
Zhishuai Zhang\textsuperscript{1},\\ Yan Wang\textsuperscript{1}, Alan Yuille\textsuperscript{1}}
\authorrunning{Y. Zhou et al.}

\institute{
\textsuperscript{1}The Johns Hopkins University\\
\textsuperscript{2}University of Maryland \& R. Adams Cowley Shock Trauma Center\\}
\maketitle              

\vspace{-1em}
\begin{abstract}
Trauma is the worldwide leading cause of death and disability in those younger than 45 years, and pelvic fractures are a major source of morbidity and mortality. Automated segmentation of multiple foci of arterial bleeding from abdominopelvic trauma CT could provide rapid objective measurements of the total extent of active bleeding, potentially augmenting outcome prediction at the point of care, while improving patient triage, allocation of appropriate resources, and time to definitive intervention. In spite of the importance of active bleeding in the quick tempo of trauma care, the task is still quite challenging due to the variable contrast, intensity, location, size, shape, and multiplicity of bleeding foci. 
Existing work presents a heuristic rule-based segmentation technique which requires multiple stages and cannot be efficiently optimized end-to-end. To this end, we present, Multi-Scale Attentional Network (MSAN), the first yet reliable end-to-end network, for automated segmentation of active hemorrhage from contrast-enhanced trauma CT scans. MSAN consists of the following components: 1) an encoder which fully integrates the global contextual information from holistic 2D slices; 2) a multi-scale strategy applied both in the training stage and the inference stage to handle the challenges induced by variation of target sizes;  3) an attentional module to further refine the deep features, leading to better segmentation quality; and
4) a multi-view mechanism to leverage the 3D information. MSAN reports a significant improvement of more than $7\%$ compared to prior arts in terms of DSC.

\end{abstract}

\section{Introduction}
\label{Introduction}

High-energy pelvic fractures, which are usually related to motor vehicle accidents, falls from height, or crush injury, are the second leading cause of death from acute physical trauma after brain injury. The mortality rate of pelvic fractures ranges from $5\%\sim15\%$, overall, increasing from $36\%$ to $54\%$ in those with hemorrhagic shock~\cite{sathy2009effect}. With the widespread availability of CT in trauma bays, the majority of patients with severe pelvic trauma admitted to level I trauma centers currently undergo an examination with contrast-enhanced trauma CT, in part to assess for foci of active bleeding, manifesting as contrast extravasation~\cite{cullinane2011eastern}.  The size of foci of contrast extravasation from bleeding vessels correlates with the need for blood transfusion, angiographic or surgical hemostatic intervention, and mortality, but reliable measurements of contrast extravasation volume cannot be derived at the point of care using manual, semi-automated, or shorthand diameter-based methods. Fully automated methods are necessary for real-time point-of-care decision making, treatment planning, and prognostication.

In this paper, we focus on volumetric segmentation of foci of active bleeding (i.e. contrast extravasation) after pelvic fractures. This task is of vital importance yet challenging for the following reasons: 1) hemorrhage gray levels vary from patient to patient, depending on a variety of factors (\emph{e.g.}, the rate of bleeding, the timing of the scan, and the patient’s physiologic state after trauma), 2) hemorrhage boundaries are often very poorly defined and highly irregular; and 3) the intensity levels are inconsistent throughout the region of a hemorrhagic focus. Prior works have utilized semi-automated threshold- or region growing-based methods using post-processing software~\cite{dreizin2018ct}. However, these techniques are too time-consuming for clinical use in the trauma radiology setting. To overcome this difficulty, a method~\cite{davuluri2012hemorrhage} was previously proposed to first utilize spatial contextual information from artery and bone to detect the hemorrhage, and then employ a rule-based strategy to refine the segmentation results. This heuristic approach requires multiple stages which cannot be efficiently optimized end-to-end. Moreover, this method cannot properly handle other challenges such as variation of target sizes and ambiguous boundaries.

\newcommand{\colheight}{2.2cm}
\begin{figure}[!tb]
\centering
% \resizebox{2.0\textwidth}{!}{
% \hspace{.5em}
% \includegraphics[width=0.248\textwidth]{Figures/bleed/f1-min.pdf}\hfill
% \includegraphics[width=0.248\textwidth]{Figures/bleed/f2-min.pdf}\hfill
% \includegraphics[width=0.248\textwidth]{Figures/bleed/f3-min.pdf}\hfill
% \includegraphics[width=0.248\textwidth]{Figures/bleed/f4-min.pdf}\hfill
% % \hspace{.5em}
% \vspace{.1em}
% % \hspace{.5em}
% \includegraphics[width=0.248\textwidth]{Figures/bleed/f5-min.pdf}\hfill
% \includegraphics[width=0.248\textwidth]{Figures/bleed/f6-min.pdf}\hfill
% \includegraphics[width=0.248\textwidth]{Figures/bleed/f7-min.pdf}\hfill
% \includegraphics[width=0.248\textwidth]{Figures/bleed/f8-min.pdf}\hfill
% % }
% % \hspace{.5em}
\includegraphics[width=0.85\linewidth]{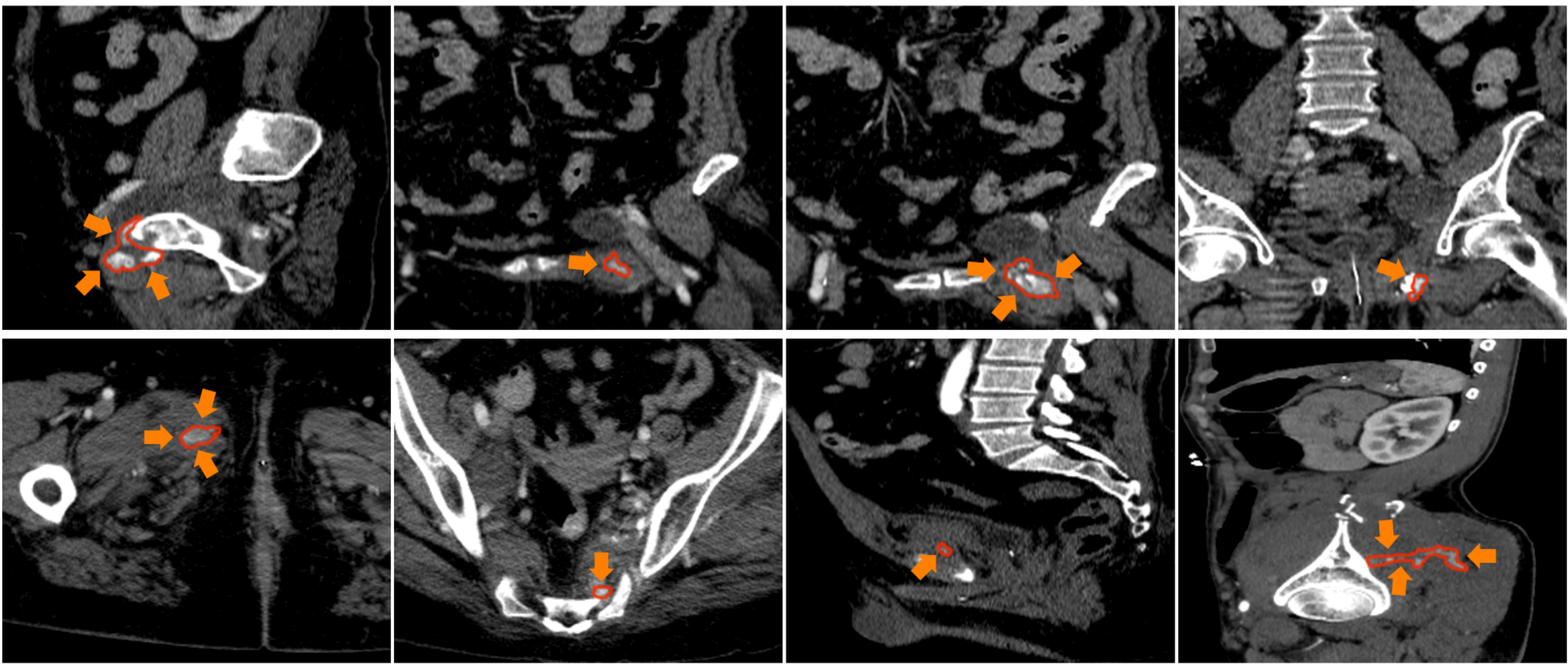}
\vspace{-1.2em}
\caption{\textbf{Visual examples of pelvic CT scans from axial/coronal/saggital views}. Red contour denotes the boundaries of the active hemorrhage, where we can observe large variations of shape and textures.}\label{fig:main}
\vspace{-1.7em}
\end{figure}

Recently, the emerge of deep learning has largely advanced the field of computer aided diagnosis (CAD). Riding on the success of convolutional neural networks, \eg, fully convolutional networks~\cite{Long_2015_Fully}, researchers have achieved accurate segmentation on many medical image analysis tasks~\cite{zhou2018semi,Ronneberger_2015_UNet,zhou2019prior,Roth_2016_Spatial,zhou2017fixed,zhu2019anatomynet}. Existing coarse-to-fine methods~\cite{zhou2017fixed,zhou2017deep,yu2018recurrent}, which propose to refine segmentation results through explicit cropping of a single region
of interest (ROI) are more suitable for single connected structures such as the pancreas or liver, while sites of active bleeding are frequently discontinuous and multi-focal and occur in widely disparate vascular territories. Herein, we present a multi-scale attentional network (MSAN), for segmenting active bleed after pelvic features, the first yet reliable framework, for segmenting active bleed after pelvic features. Specifically, our framework is able to 1) fully exploit contextual information from holistic 2D slices via using an encoder which is capable of extracting the global contextual information across different levels of image features; 2) efficiently handle the variation of active hemorrhage sizes by adopting multi-scale strategies during the training phase and the testing phase; 3) deal with the ambiguous boundaries by utilizing an attentional mechanism to better enhance the discrimination between trauma region and non-trauma region; 4) utilize the aggregation of multiple views (\ie, Coronal, Sagittal and Axial views)  to further leverage the 3D information. To assess the effectiveness of our framework, we collect a dataset of 65 patients with pelvic fractures and active hemorrhage with widely varying degrees of severity. For each case, every pixel/voxel of active hemorrhage was manually labeled by an experienced radiologist. Unlike the  previously described heuristic method which used crude and not widely adopted measurements of accuracy such as missegmented area~\cite{davuluri2012hemorrhage}, we employed the Dice-S{\o}rensen coefficient (DSC) for evaluation based on pixel/voxel-wise predictions. Experimental results demonstrate the superiority of our framework compared with a series of 2D/3D state-of-the-art deep learning algorithms.

\section{Multi-Scale Attentional Network}
\label{Approach}

\subsection{Overall Framework}
\label{sec:overview}
We denote a 3D CT-scanned image as $\mathbf{X}$ with size $W\times H\times L$, where each element of $\mathbf{X}$ indicated the Housefield Unit (HU) of a voxel.
The corresponding binary ground-truth segmentation mask is denoted as $\mathbf{Y}$ where ${y_i}={1}$ indicates a foreground voxel.
Consider a segmentation model $\mathbb{M}:{\mathbf{Z}}={\mathbf{f}\!\left(\mathbf{X};\boldsymbol{\Theta}\right)}$,
where $\mathbb{M}$ is parameterized by  $\boldsymbol{\Theta}$, our goal is to predict a binary output volume $\mathbf{Z}$ of the same dimension as $\mathbf{X}$.
We denote $\mathcal{Y}$ and $\mathcal{Z}$ as the set of foreground voxels in the ground-truth and prediction,
{\ie}, ${\mathcal{Y}}={\left\{i\mid y_i=1\right\}}$ and ${\mathcal{Z}}={\left\{i\mid z_i=1\right\}}$.
The accuracy of segmentation is evaluated by the Dice-S{\o}rensen coefficient (DSC):
${\mathrm{DSC}\!\left(\mathcal{Y},\mathcal{Z}\right)}=
    {\frac{2\times\left|\mathcal{Y}\cap\mathcal{Z}\right|}{\left|\mathcal{Y}\right|+\left|\mathcal{Z}\right|}}$.
This metric falls in the range of $\left[0,1\right]$, and DSC = 1 implies a perfect segmentation.

Following~\cite{zhou2017fixed,yu2018recurrent,Roth_2016_Spatial}, 3 sets of images, {\ie}, $\mathbf{X}_{\mathrm{C},w}$ (${w}={1,2,\ldots,W}$),
$\mathbf{X}_{\mathrm{S},h}$ (${h}={1,2,\ldots,H}$) and $\mathbf{X}_{\mathrm{A},l}$ (${l}={1,2,\ldots,L}$) are obtained along three axes. 
The subscripts $\mathrm{C}$, $\mathrm{S}$ and $\mathrm{A}$ stand for ``coronal'', ``sagittal'' and ``axial'', respectively. We train an individual model $\mathbb{M}$ for each of the three viewpoints.
Without loss of generality, we consider a 2D slice along the {\em axial} view, denoted by $\mathbf{X}_{\mathrm{A},l}$.
Our goal is to infer a binary segmentation mask $\mathbf{Z}_{\mathrm{A},l}$ of the same dimensionality.
In the context of deep networks~\cite{Long_2015_Fully,Chen_2015_Semantic},
it is achieved by computing a {\em probability map}
${\mathbf{P}_{\mathrm{A},l}}={\mathbf{f}\!\left[\mathbf{X}_{\mathrm{A},l};\boldsymbol{\theta}\right]}$,
where $\mathbf{f}\!\left[\cdot;\boldsymbol{\theta}\right]$ is the architecture as in Fig.~\ref{fig:aspp}(a). This network contains an encoder (Sec.~\ref{sec:baseline}) to extract different levels of features for distilling global context and an attentional module (Sec.~\ref{sec:Attention module}) as further refinement.

Specifically, we apply Atrous Spatial Pyramid Pooling  (ASPP)~\cite{Chen_2015_Semantic} at the end of the backbone model to extract high-level features with enriched global context.
Meanwhile, the low-level features extracted from earlier layers which contain local information are fed to an attentional module to distill more useful information.
The refined low-level features are then concatenated with high-level features extracted by ASPP and fed to the final classifier layer, which outputs probabilities $\mathbf{P}_{\mathrm{A},l}$, $\mathbf{P}_{\mathrm{C},l}$ and $\mathbf{P}_{\mathrm{S},l}$ which are then binarized into $\mathbf{Z}_{\mathrm{A},l}$, $\mathbf{Z}_{\mathrm{C},l}$ and $\mathbf{Z}_{\mathrm{S},l}$ respectively.
The final segmentation outcome can be fused from the three views via majority voting~\cite{zhou2017fixed,yu2018recurrent}. Multi-scale processing~\cite{Kamnitsas_2016_Efficient,Chen_2015_Semantic} is used in both the training stage and the inference stage to further enhance the segmentation accuracy, especially for small targets. As illustrated in Fig.~\ref{fig:aspp}, different rescaled version of the original image are fed to the network during training. During the testing stage, to produce the final segmentation mask, the output from different scales are fused by taking at each position the average response. If the average probability is larger than a certain threshold $\rho$ it is regarded as foreground otherwise it is regarded as background.

\subsection{Encoder Backbone Architecture}
\label{sec:baseline}
Atrous Convolution has been widely applied in computer vision problems, which can efficiently allow for larger receptive field via controlling atrous rates. Given an input feature map $\bm{x}$, atrous convolution is applied over $\bm{x}$ as follows:

\begin{equation}
  \bm{y}[\bm{i}] = \sum_{\bm{k}} \bm{x}[\bm{i} + r \cdot \bm{k}] \bm{w}[\bm{k}],
\end{equation}
where $\bm{i}$ and $\bm{w}$ denote the spatial 
location and the convolution filter, respectively. $r$ stands for the atrous rate. 

\begin{figure}[t]
\centering
\includegraphics[width=0.85\linewidth]{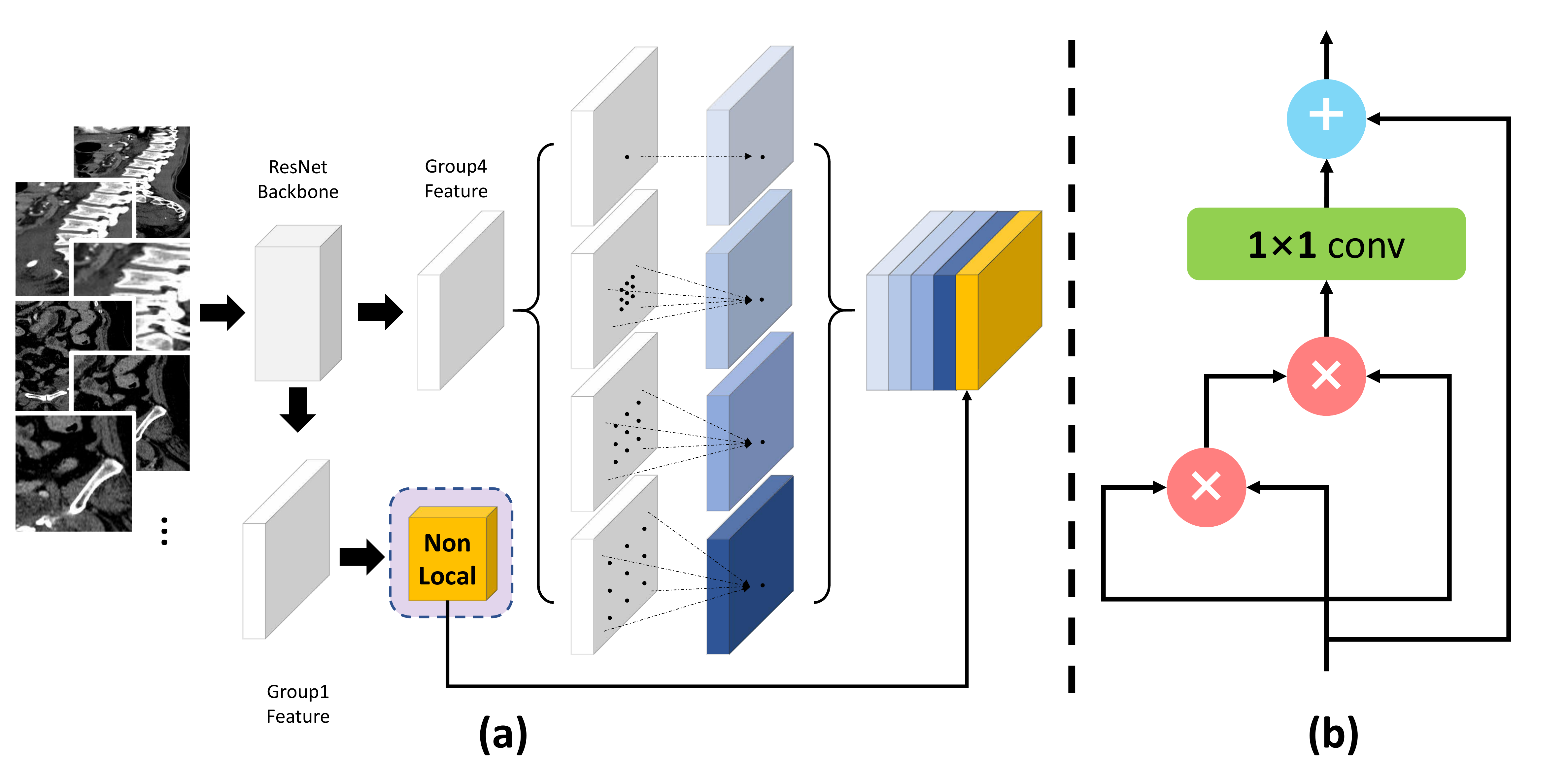}
\vspace{-1.2em}
\caption{(a) The network architecture structure of MSAN.  Low-level features are refined by an attentional module. Meanwhile ASPP is applied at the end of the backbone model to extract high-level features with enriched global context. (b) Our implementation of the attentional module, where we use nonlocal means~\cite{Wang2018} as the main operation.}
\vspace{-1.3em}
\label{fig:aspp}
\end{figure}

Atrous Spatial Pyramid Pooling (ASPP) is originated from Spatial Pyramid Pooling~\cite{he2015spatial}. The main difference is that ASPP uses atrous convolution which allows for larger field-of-view during training and thus can efficiently integrate global contextual information. As a strong contextual aggregation module~\cite{Chen_2015_Semantic}, ASPP is applied (see Fig.~\ref{fig:aspp}(a)) so that the contextual information from artery and bone can be better exploited. In our experiment, we set the atrous rates to be $\{12, 24, 36\}$, respectively. 

\subsection{Attentional Module}
\label{sec:Attention module}
We adapt the non-local block \cite{Wang2018} as the attentional module in our framework. Specifically, it first computes an attention map $y$ of an input feature map $x$ by taking a weighted average of features in all spatial locations $\mathcal{L}$:
\begin{equation}
\label{eq:nonlocal}
\bm{y_i} = \frac{1}{\mathcal{C}(\bm{x})} \sum_{\forall \bm{j} \in \mathcal{L}} f(\bm{x_i}, \bm{x_j})\cdot \bm{x_j},
\end{equation}
where i and j are spatial indices. A pairwise function $f(\bm{x_i}, \bm{x_j})$ is used to compute the spatial attention coefficients between each i and all j. And these coefficients are applied as the weighting of the input feature to better prune out irrelevant background features and thereby distinguish salient image regions.
$\mathcal{C}(\bm{x})$ is a normalization function. We use the dot product version in \cite{Wang2018} by setting $f(\bm{x_i}, \bm{x_j}) = \bm{x_i}^\text{T} \bm{x_j}$ and $\mathcal{C}(\bm{x}) \!=\! N$, where $N$ is the number of pixels in $\bm{x}$. 

Following \cite{Wang2018}, the attention map $\bm{y}$ is then processed by a 1$\times$1 convolutional layer and added to the input feature map $\bm{x}$ to obtain the final output $\bm{z}$, \ie, $\bm{z} = \bm{w y} + \bm{x}$, where $\bm{w}$ is the weight of the convolutional layer. An illustration our our attentional module can be found in Fig.~\ref{fig:aspp}(b).

\section{Experiments}
\label{Experiments}

\subsection{Dataset and Evaluation}
\label{Experiments:DatasetEvalutation}
We have collected 65 studies were routinely acquired with 64 section or higher MDCT scanners in the trauma bay in either the late arterial or portal venous phase of enhancement. We use 45 cases for training and evaluate the segmentation performance on the rest 20 cases. Note that ~\cite{davuluri2012hemorrhage} was studied on only 12 cases, which, to the best of our knowledge, was the first and only curated dataset with manual ground truth label masks. Therefore our dataset can be considered as a valid set for evaluation.
The metric we use is DSC, which measures 
the similarity between the prediction voxel set $\mathcal{Z}$ and the ground-truth set $\mathcal{Y}$,
with the mathematical form of ${\mathrm{DSC}\!\left(\mathcal{Z},\mathcal{Y}\right)}=
    {\frac{2\times\left|\mathcal{Z}\cap\mathcal{Y}\right|}{\left|\mathcal{Z}\right|+\left|\mathcal{Y}\right|}}$.
\vspace{-0.4cm}
\subsection{Implementation details}
\label{sec:imp_details}
Our implementations are based on Tensorflow. We used two standard architectures, \emph{i.e.}, ResNet-50 and ResNet-101~\cite{He_2016_Deep} as backbone models.
All our segmentation experiments were performed on the whole pelvic CT scan and were run on Tesla V100 GPU. For data pre-processing, following~\cite{Roth_2016_Spatial}, we simply truncated the raw intensity values to be within the range of $[-80, 320]$~HU and then normalized each raw CT case to $[0, 255.0]$. Random rotation of $[0,15]$ is used as online data augmentation. A \emph{poly} learning policy is applied with an initial learning rate of $0.05$ with a decay power of 0.9. We follow  \cite{zhou2017fixed,Roth_2016_Spatial,yu2018recurrent} to use ImageNet pretrained model for initialization.
% which has been demonstrated to be useful in various studies

\subsection{Results and Discussions}
\label{ablation: Results}
\begin{table}[t]
\centering
\scriptsize
\setlength{\tabcolsep}{2mm}
\renewcommand\arraystretch{1}
\resizebox{0.9\linewidth}{!}{
\begin{tabular}{l|c |c| c | c  | c}
\hline
\hline
Model   &scale=1.0  & scale=1.25   &scale=1.5 &scale=1.75 & Avg. Dice  \\
\hline
ResNet50-single-scale			&\checkmark & -   & -  &-   &35.96\% \\
ResNet50-single-scale		& - &\checkmark &- &-    & 48.14\% \\
ResNet50-single-scale			&- &- &\checkmark  &-   & 47.71\% \\
ResNet50-single-scale	&- &- &- &\checkmark  & 46.29\%   \\
\hline
ResNet50-2-scale	     &- &\checkmark &\checkmark &-  & 52.75\%   \\
ResNet50-MSAN-2-scale	    &- &\checkmark &\checkmark &-  & 54.31\%   \\
ResNet50-3-scale	    &- &\checkmark &\checkmark &\checkmark & 54.40\%   \\
ResNet50-MSAN-3-scale	     &- &\checkmark &\checkmark &\checkmark & 55.61\%   \\
\hline
\hline
ResNet101-single-scale			&\checkmark & -   & -  &-   & 37.53\% \\
ResNet101-single-scale		& - &\checkmark &- &-    & 46.38\% \\
ResNet101-single-scale			&- &- &\checkmark  &-   & 52.67\% \\
ResNet101-single-scale	&- &- &- &\checkmark  & 54.56\%   \\
\hline
ResNet101-2-scale	     &- &\checkmark &\checkmark &-  & 54.98\%   \\
ResNet101-MSAN-2-scale	    &- &\checkmark &\checkmark &-  & 55.70\%   \\
ResNet101-3-scale	    &- &\checkmark &\checkmark &\checkmark & 58.72\%   \\
ResNet101-MSAN-3-scale		     &- &\checkmark &\checkmark &\checkmark &\textbf{59.89\%}   \\
\hline
Zhou \etal~\cite{zhou2017fixed}	     &- &- &- &- & 50.15\%   \\
Yu \etal~\cite{yu2018recurrent}		     &- &- &- &- & 52.12\%   \\
3D-UNet~\cite{cciccek20163d}	     &- &- &- &- & 40.81\%   \\
\hline
\hline
\end{tabular}
}
%\end{adjustbox}
\caption{DSC comparison of active bleed segmentation. ResNet101-MSAN-3-scale achieves the best performance of 59.89\%, surpassing the prior art by more than $7\%$.}
\label{tbl:results}
\vspace{-2.7em}
\end{table}

All results are summarized in Table~\ref{tbl:results}, where we list thorough comparisons under different configuration of network architecture (\emph{i.e.}, ResNet50 and ResNet101~\cite{He_2016_Deep}) and scales (\emph{i.e.}, $scales=\{1.0, 1.25, 1.5, 1.75\}$). Note that we use larger scales ($\geq1.0$) since our goal is to segment small targets. Under different settings, our method consistently outperforms others, indicating the effectiveness of MSAN. 

\paragraph{Efficacy of multi-scale processing.}
As shown in Table~\ref{tbl:results}, larger scales generally lead to better results. For instance, using ResNet50 as the backbone model, the performance under $scale=1.0$ is $\sim10\%$ lower than that under other larger scales. ResNet101-single-scale yields the best result of $54.56\%$ under $scale=1.75$, which is more than $17\%$ better than using the scale of $1.0$. These facts all indicate the efficacy of utilizing larger scales. Another observation is that the integration of more scales also leads to better segmentation quality than using just one scale. Using either ResNet50 or ResNet101 as the backbone, 3-scales always yield better results than 2-scales/single-scale, which shows that the learned knowledge from these different scales is complementary to each other. Therefore combining the information from these different scales can be beneficial for handling targets with a large variety of sizes, such as active bleed in our study.

\begin{figure}[tb]
\centering
\includegraphics[width=0.8\linewidth]{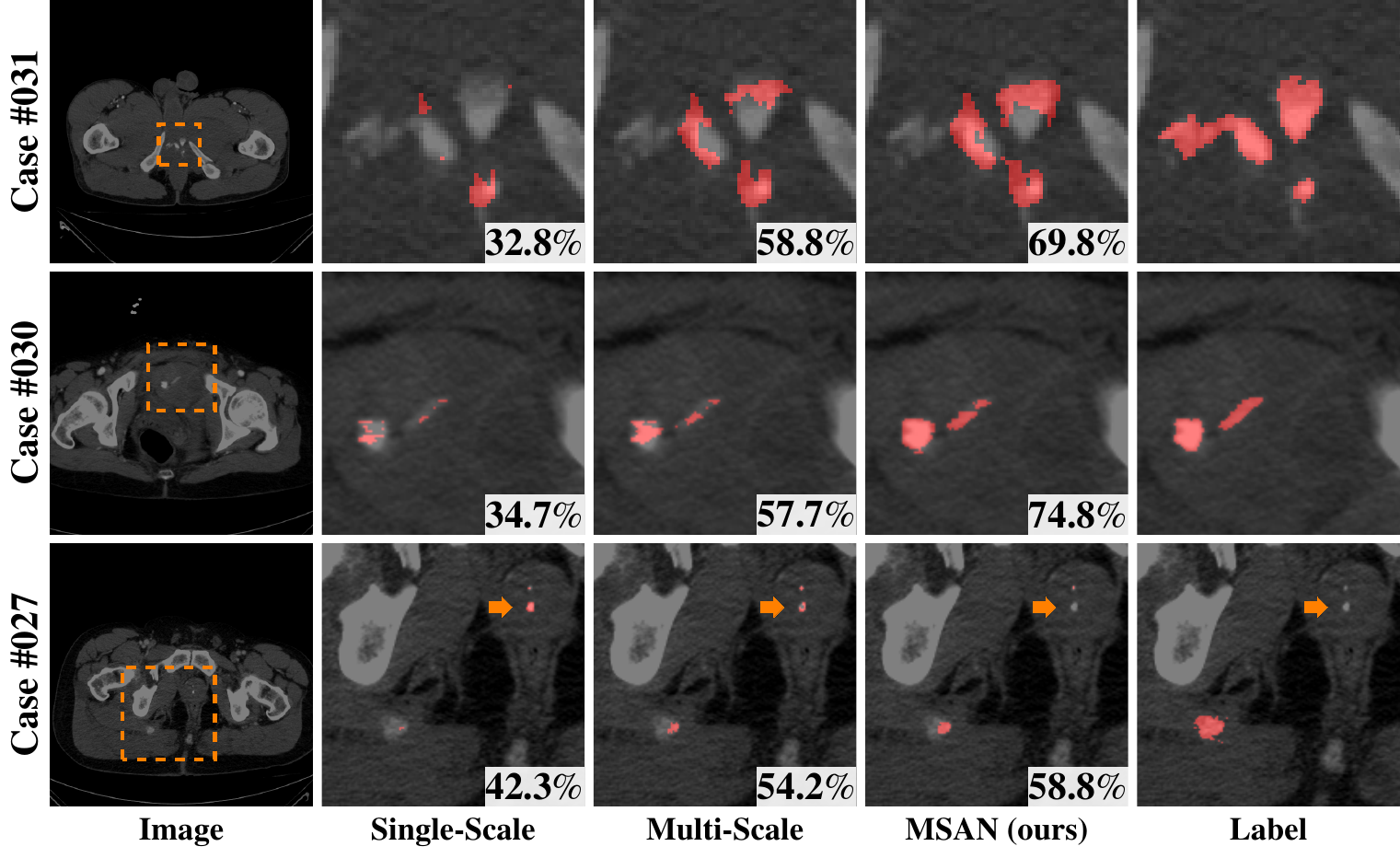}
\vspace{-1.5em}
\caption{\textbf{Qualitative comparison of different methods.} from left to right: original CT image, predictions of single-scale method ($scale=1.50$), multi-scale method ($scale=\{1.25, 1.50, 1.75\}$), MSAN and the manual label. (Best viewed in color)}
\label{fig:qualitative}
\vspace{-1.3em}
\end{figure}

\paragraph{Efficacy of the attentional module.}
Meanwhile, we also witness additional benefit from the attentional module. For instance, ResNet50-MSAN-3-scale observes an improvement of $1.17\%$ compared with ResNet101-3-scale; ResNet101-MSAN-2-scale) observes an improvement of $0.72\%$ compared with ResNet101-2-scale. A similar improvement can be also witnessed for ResNet-50. Three qualitative examples are shown in Fig.~\ref{fig:qualitative}, where MSAN consistently outperforms other existing methods. For case 027, our MSAN successfully removes the outlier (indicated by the orange arrows) which is detected as false positives by other methods. This further justifies that the usage of attentional mechanisms can indeed refine the results and diminish non-trauma outliers.

Overall, our proposed MSAN observes a significant performance gain under different settings, which shows the generality and soundness of our approach. 
Additionally, we also compare our method with other state-of-art 3D segmentation methods including~\cite{zhou2017fixed},~\cite{yu2018recurrent} and 
% \footnote{implementation based on https://github.com/DLTK/DLTK}
~\cite{cciccek20163d}.
Our method outperforms all these methods significantly (\emph{p-value}s for testing significant difference satisfy $p < 0.0001$), which further demonstrates the effectiveness of our approach. 
In order to further validate the generality and stability of MSAN, we directly test on a newly collected additional 15 cases without any retraining. Our method obtains an average DSC of $50.19\%$, whereas prior arts report $44.15\%$ (\cite{yu2018recurrent}), $35.14\%$ (\cite{zhou2017fixed}) and $27.32\%$ (\cite{cciccek20163d}). MSAN significantly outperforms these methods.

\section{Conclusions}
\label{Conclusions}
In this paper, we present Multi-Scale Attentional Network (MSAN), an end-to-end framework for automated segmentation of active hemorrhage from pelvic CT scans. Our proposed MSAN substantially improves the segmentation accuracy by more than $7\%$ compared with prior arts. We note this framework can be practical in assisting radiologists
for clinical applications, since the annotation in 3D volumes requires massive labor from radiologists.

\vspace{0.25em}
{\footnotesize \noindent {\bf Acknowledgements.}
This work was supported by NIBIB (National Institute of Biomedical Imaging and Bioengineering)/NIH under award number K08EB027141, University of Maryland Institute for Clinical and Translational Research Accelerated Translational Incubator Pilot (ATIP) award and Radiologic Society of North America (RSNA) Research Scholar Award \#1605. We thank Fengze Liu, Yingda Xia, Qihang Yu and Zhuotun Zhu for instructive discussions.}
\bibliographystyle{splncs04}
\bibliography{paper}
\end{document}